\documentclass[twocolumn]{emulateapj}
\usepackage{graphicx}
\usepackage{natbib}
\usepackage{subfigure}
\usepackage{url}
\usepackage{verbatim}
\usepackage{ amssymb }
\usepackage{amsmath}
\begin{document}

\title{Mining Planet Search Data for Binary Stars: The $\psi^1$ Draconis system}
\author{Kevin Gullikson \altaffilmark{1}}
\author{Michael Endl \altaffilmark{1}}
\author{William D. Cochran \altaffilmark{1}}
\author{Phillip J. MacQueen \altaffilmark{1}}

\altaffiltext{1}{University of Texas, Astronomy Department. 2515 Speedway, Stop C1400. Austin, TX 78712 \email{kgulliks@astro.as.utexas.edu}}

\begin{abstract}
Several planet-search groups have acquired a great deal of data in the form of time-series spectra of several hundred nearby stars with time baselines of over a decade. While binary star detections are generally not the goal of these long-term monitoring efforts, the binary stars hiding in existing planet search data are precisely the type that are too close to the primary star to detect with imaging or interferometry techniques. We use a cross-correlation analysis to detect the spectral lines of a new low-mass companion to $\psi^1$ Draconis A, which has a known roughly equal-mass companion at ${\sim}680$ AU. We measure the mass of $\psi^1$ Draconis C as $M_2 = 0.70 \pm 0.07 M_{\odot}$, with an orbital period of ${\sim}20$ years. This technique could be used to characterize binary companions to many stars that show large-amplitude modulation or linear trends in radial velocity data.
\end{abstract}

\maketitle

\section{Introduction}
\label{sec:intro}
Several groups \citep[e.g.][]{Wittenmyer2006, Fischer2009, Pepe2011} have used the radial velocity method to search for planets around nearby stars for well over a decade, and have collectively uncovered several hundred planets to date. Close binary stars are usually cut from the star sample because they complicate the detection method \citep[e.g.][]{Bergmann2015}, and because they have long been suspected to inhibit planet formation by quickly destroying \citep{Kraus2012} or depleting \citep{Harris2012} the planet-forming disk.

Previously unknown stellar binary companions are nonetheless still uncovered in planet--search data through large--amplitude linear trends or even full long-period orbits, but may be ignored since the goal is to find planet--mass companions. Since binary stars are usually excluded in the star sample, companions that are found tend to have extreme flux- and mass-ratios. Binary stars with extreme mass-ratios on orbits with $\sim 10$ year timescales are precisely the ones that are most difficult to detect and characterize with imaging techniques, and so they should not be ignored.

Several groups have recently worked towards using high-resolution spectroscopy to search for very faint companions to nearby stars, both in the context of detecting emission \citep{Snellen2010, Gullikson2013} or reflection \citep{Martins2013} from ``Hot Jupiter'' planets, and in the context of detecting stellar binary systems with high contrast ratios \citep[e.g.][]{Gullikson2013_2, Kolbl2015}. Those groups all use a cross-correlation analysis to search directly for the spectral lines of the faint companion and mainly differ in their treatment of the primary star and telluric lines.

In this paper we use data from the McDonald Observatory Planet Search team to examine the $\psi^1$ Draconis system, which consists of an F5IV--V star ($\psi^1$ Dra A) orbited by a G0V star ($\psi^1$ Dra B) with angular separation $30''$ \citep{WDS}. \cite{Tokovinin2002} searched for signs of a spectroscopic companion to $\psi^1$ Dra A from 1991--1995, but found no radial velocity variation. More recently \cite{Toyota2009} noted a linear trend in their radial velocity measurements, and predicted a companion with $M > 50 M_J$. Our data have a much longer time baseline than either of the previous studies, and show a significant fraction of the orbit which has recently reached quadrature. Furthermore, \citet{Endl2015} use adaptive-optics imaging to detect a $\sim 4500\ K$ companion 155 mas from $\psi^1$ Dra A, which they hypothesize is the source of the orbital motion seen in the primary-star radial-velocity measurements. 

Here, we use all of our spectra of $\psi^1$ Dra A to search directly for the spectral lines of the companion and measure the system mass ratio. We describe the observations and data reduction in Section \ref{sec:obs}, and the method we use to search for the companion in Section \ref{sec:method}. Finally, we estimate the mass-ratio of the system and give the parameters for the companion in Section \ref{sec:orbit}.

\section{Observations and Data Reduction}
\label{sec:obs}
All data were taken at the 2.7 m Harlan J. Smith Telescope at McDonald observatory using the 2dcoud\'e \'echelle spectrograph \citep{TS23} at a resolving power $R\equiv \frac{\lambda}{\Delta \lambda} = 60000$. The starlight was filtered through a temperature-stabilized $I_2$ cell to imprint many sharp absorption lines on each spectrum to use for both a precise velocity metric \citep{Butler1996} and to model the instrument profile \citep{Endl2000}. The raw CCD data were reduced with standard IRAF\footnote{IRAF is distributed by the National Optical Astronomy Observatories, which are operated by the Association of Universities for Research in Astronomy, Inc., under cooperative agreement with the National Science Foundation.} tasks, and include steps for overscan trimming, bad--pixel processing, bias--frame subtraction, scattered--light removal, flat--field division, order extraction, and wavelength solution fitting using a Th--Ar calibration lamp spectrum. Particularly strong cosmic--ray hits were removed manually by interpolating across nearby pixels.

We used the \emph{Austral} code \citep{Endl2000} to measure the differential radial velocity of $\psi^1$ Dra A at each observation by comparing each spectrum to a high signal-to-noise ratio template spectrum of the same star. We provide the raw velocity measurements in Table \ref{tab:rv_data}, as well as the velocities shifted into the system velocity rest frame. The velocity shift necessary to convert from the differential radial velocities to that frame is found in Section \ref{sec:orbit}. Table \ref{tab:rv_data} also gives the measurements of the companion radial velocity (described in the next section).

\begin{deluxetable}{lrrrrrr}
\tabletypesize{\footnotesize}
\tablewidth{0pt}
\tablecaption{Observations of $\psi^1$ Dra A \label{tab:rv_data}}
\tablehead{
\colhead{Julian Date} & \multicolumn{3}{c}{Primary RV (km/s)} & \multicolumn{3}{c}{Secondary RV (km/s)} \\
& \colhead{raw} & \colhead{shifted} & \colhead{$\sigma$}  & \colhead{raw} & \colhead{shifted}  & \colhead{$\sigma$} }

\startdata

 2451809.66 &   1.927 &  -2.174 &   0.013 &  \nodata &  \nodata &  \nodata \\
 2451809.67 &   1.929 &  -2.172 &   0.014 &  \nodata &  \nodata &  \nodata \\
 2452142.68 &   1.841 &  -2.259 &   0.012 &  \nodata &  \nodata &  \nodata \\
 2453319.64 &   2.433 &  -1.668 &   0.011 &  \nodata &  \nodata &  \nodata \\
 2453585.85 &   2.559 &  -1.542 &   0.010 &  \nodata &  \nodata &  \nodata \\
 2453585.88 &   2.550 &  -1.551 &   0.011 &  \nodata &  \nodata &  \nodata \\
 2453634.64 &   2.654 &  -1.446 &   0.011 &  \nodata &  \nodata &  \nodata \\
 2453635.62 &   2.554 &  -1.547 &   0.009 &  \nodata &  \nodata &  \nodata \\
 2453655.64 &   2.711 &  -1.390 &   0.009 &  \nodata &  \nodata &  \nodata \\
 2453655.64 &   2.780 &  -1.321 &   0.027 &  \nodata &  \nodata &  \nodata \\
 2453689.54 &   2.665 &  -1.436 &   0.008 &  \nodata &  \nodata &  \nodata \\
 2453907.85 &   2.960 &  -1.141 &   0.011 &  \nodata &  \nodata &  \nodata \\
 2453928.80 &   2.858 &  -1.243 &   0.012 &  \nodata &  \nodata &  \nodata \\
 2454019.60 &   2.930 &  -1.171 &   0.012 &  \nodata &  \nodata &  \nodata \\
 2454279.75 &   3.068 &  -1.033 &   0.011 &  \nodata &  \nodata &  \nodata \\
 2454279.76 &   3.056 &  -1.044 &   0.010 &  \nodata &  \nodata &  \nodata \\
 2454309.79 &   3.021 &  -1.080 &   0.013 &  \nodata &  \nodata &  \nodata \\
 2454345.63 &   3.270 &  -0.830 &   0.010 &  \nodata &  \nodata &  \nodata \\
 2454401.56 &   3.155 &  -0.945 &   0.009 &  \nodata &  \nodata &  \nodata \\
 2454662.93 &   3.349 &  -0.752 &   0.015 &    -4.34 &     0.36 &     0.37 \\
 2454665.77 &   3.486 &  -0.615 &   0.014 &    -3.95 &     0.75 &     0.35 \\
 2454665.77 &   3.492 &  -0.609 &   0.015 &    -4.06 &     0.64 &     0.36 \\
 2454730.71 &   3.457 &  -0.644 &   0.014 &    -4.09 &     0.68 &     0.36 \\
 2455100.57 &   3.875 &  -0.226 &   0.016 &    -5.14 &     0.04 &     0.44 \\
 2455100.58 &   3.891 &  -0.210 &   0.014 &    -5.02 &     0.16 &     0.44 \\
 2455398.75 &   4.209 &   0.108 &   0.015 &    -6.49 &    -0.89 &     0.51 \\
 2455790.72 &   4.977 &   0.876 &   0.021 &    -8.65 &    -2.34 &     0.67 \\
 2455869.58 &   5.211 &   1.111 &   0.017 &    -8.14 &    -1.66 &     0.60 \\
 2455910.57 &   5.321 &   1.221 &   0.018 &    -8.40 &    -1.82 &     0.63 \\
 2455992.02 &   5.538 &   1.437 &   0.012 &    -9.01 &    -2.22 &     0.65 \\
 2456016.93 &   5.659 &   1.558 &   0.014 &    -9.23 &    -2.37 &     0.62 \\
 2456106.78 &   5.784 &   1.683 &   0.015 &   -10.36 &    -3.24 &     0.67 \\
 2456138.84 &   5.944 &   1.844 &   0.020 &   -10.73 &    -3.51 &     0.73 \\
 2456145.65 &   5.947 &   1.846 &   0.025 &   -11.38 &    -4.14 &     0.80 \\
 2456145.66 &   5.929 &   1.828 &   0.018 &   -11.43 &    -4.19 &     0.79 \\
 2456145.66 &   5.965 &   1.864 &   0.018 &   -11.15 &    -3.91 &     0.79 \\
 2456173.73 &   5.955 &   1.854 &   0.018 &   -10.82 &    -3.48 &     0.75 \\
 2456401.97 &   6.964 &   2.864 &   0.014 &   -14.11 &    -5.87 &     0.69 \\
 
\enddata

\end{deluxetable}

\begin{deluxetable}{lrrrrrr}
\tabletypesize{\footnotesize}
\tablewidth{0pt}
\tablenum{1}
\tablecaption{Observations of $\psi^1$ Dra A (continued)}
\tablehead{
\colhead{Julian Date} & \multicolumn{3}{c}{Primary RV (km/s)} & \multicolumn{3}{c}{Secondary RV (km/s)} \\
& \colhead{raw} & \colhead{shifted} & \colhead{$\sigma$}  & \colhead{raw} & \colhead{shifted}  & \colhead{$\sigma$} }

\startdata

 2456401.97 &   6.941 &   2.841 &   0.012 &   -14.39 &    -6.15 &     0.68 \\
 2456433.74 &   7.238 &   3.138 &   0.013 &   -14.54 &    -6.14 &     0.62 \\
 2456433.74 &   7.209 &   3.108 &   0.012 &   -14.61 &    -6.21 &     0.65 \\
 2456435.87 &   7.208 &   3.108 &   0.015 &   -14.73 &    -6.33 &     0.64 \\
 2456435.87 &   7.205 &   3.104 &   0.015 &   -15.08 &    -6.68 &     0.64 \\
 2456461.87 &   7.358 &   3.257 &   0.012 &   -14.96 &    -6.42 &     0.63 \\
 2456461.88 &   7.351 &   3.250 &   0.015 &   -14.65 &    -6.11 &     0.65 \\
 2456461.88 &   7.326 &   3.225 &   0.016 &   -14.60 &    -6.06 &     0.61 \\
 2456465.80 &   7.297 &   3.196 &   0.014 &   -14.74 &    -6.18 &     0.53 \\
 2456497.86 &   7.574 &   3.473 &   0.019 &   -15.86 &    -7.13 &     0.73 \\
 2456519.62 &   7.765 &   3.664 &   0.015 &   -16.78 &    -7.93 &     0.59 \\
 2456525.66 &   7.725 &   3.624 &   0.017 &   -16.27 &    -7.38 &     0.64 \\
 2456560.58 &   7.812 &   3.711 &   0.013 &   -16.32 &    -7.22 &     0.90 \\
 2456564.59 &   7.781 &   3.680 &   0.015 &   -16.18 &    -7.05 &     0.86 \\
 2456613.55 &   8.089 &   3.988 &   0.016 &   -15.96 &    -6.51 &     0.91 \\
 2456614.58 &   8.139 &   4.038 &   0.012 &   -16.52 &    -7.06 &     0.85 \\
 2456755.98 &   9.308 &   5.208 &   0.014 &   -21.44 &   -10.84 &     0.74 \\
 2456759.97 &   9.366 &   5.265 &   0.015 &   -21.91 &   -11.28 &     0.76 \\
 2456784.84 &   9.603 &   5.502 &   0.017 &   -22.56 &   -11.69 &     0.81 \\
 2456816.67 &   9.895 &   5.794 &   0.014 &   -23.36 &   -12.17 &     0.74 \\
 2456816.67 &   9.907 &   5.806 &   0.015 &   -23.68 &   -12.49 &     0.72 \\
 2456860.73 &  10.402 &   6.301 &   0.016 &   -25.64 &   -13.97 &     0.88 \\
 2456860.73 &  10.421 &   6.321 &   0.015 &   -25.77 &   -14.11 &     0.83 \\
 2456885.62 &  10.607 &   6.507 &   0.015 &   -27.23 &   -15.29 &     0.84 \\
 2456938.63 &  11.189 &   7.089 &   0.016 &   -29.30 &   -16.77 &     1.03 \\
 2456938.64 &  11.173 &   7.072 &   0.015 &   -28.99 &   -16.45 &     0.96 \\
 2457092.02 &  12.114 &   8.013 &   0.015 &   -31.57 &   -18.09 &     1.00 \\
 2457109.85 &  12.077 &   7.976 &   0.015 &   -32.79 &   -19.40 &     1.10 \\
 2457118.96 &  11.987 &   7.886 &   0.016 &   -31.14 &   -17.82 &     0.90 \\
 2457150.92 &  11.685 &   7.584 &   0.017 &  \nodata &  \nodata &  \nodata \\
 2457174.96 &  11.267 &   7.167 &   0.017 &   -28.58 &   -16.13 &     0.81 \\
 2457214.83 &  10.240 &   6.139 &   0.017 &   -24.68 &   -13.16 &     0.82 \\
 2457214.84 &  10.253 &   6.152 &   0.016 &   -25.36 &   -13.85 &     0.90 \\
 2457216.73 &  10.220 &   6.119 &   0.016 &   -25.13 &   -13.66 &     0.86 \\
 2457216.73 &  10.228 &   6.128 &   0.015 &   -25.15 &   -13.69 &     0.80 \\
 2457245.60 &   9.302 &   5.201 &   0.016 &   -22.21 &   -11.52 &     0.77 \\
 2457245.61 &   9.299 &   5.199 &   0.016 &   -21.85 &   -11.16 &     0.72 \\
 2457248.61 &   9.338 &   5.237 &   0.017 &   -21.66 &   -11.05 &     0.70 \\
\enddata

\tablecomments{The velocities in the `raw' columns are our actual measurements. Those in the `shifted' columns are shifted into the system velocity rest frame using the results of the orbital fit described in Section \ref{sec:orbit}.}
\end{deluxetable}

\section{Companion Search}
\label{sec:method}

We use a cross-correlation analysis inspired by recent work attempting to detect light from planetary companions around late-type stars \citep{Gullikson2013, Martins2013} to search for the companion ($\psi^1$ Dra C). We start by dividing all spectra by the blaze function of the spectrograph, and further divide them by an empirical $I_2$ cell absorption spectrum in the spectral orders with $500 < \lambda < 640$ nm. The blaze function is derived by fitting a high-order polynomial to the extracted spectrum of an incandescent light source (a flat lamp), and the empirical $I_2$ spectrum is the spectrum of a flat lamp with the $I_2$ cell inserted in the light path. Both the flat lamp and $I_2$ spectra are observed each day of each observing run. We use the Telfit code \citep{Gullikson2014} to fit and remove the unsaturated telluric absorption lines in the spectrum, and cross-correlate each residual spectrum against a Phoenix model spectrum \citep{Husser2013} with parameters

\begin{itemize}
\item $T_{\rm eff} = 4400$ K
\item $\log{g} = 4.5$ (cgs units)
\item {[}Fe/H{]} = 0.0
\end{itemize}

The model temperature was chosen on the basis of high--contrast imaging in \citet{Endl2015}, which finds a companion with approximately that temperature. We shift each CCF so that the dominant peak, which signifies the match of the M-star model template with the F-type primary star, falls at $v=0$. That effectively puts the cross-correlation functions in the rest frame of the primary star, although there is a constant velocity offset caused by small errors in the vacuum to air wavelength conversion and spectrograph wavelength drift throughout the night. We denote this shift as $\Delta v_2$ in later sections of this paper.

\begin{figure}
  \centering
  \includegraphics[width=\columnwidth]{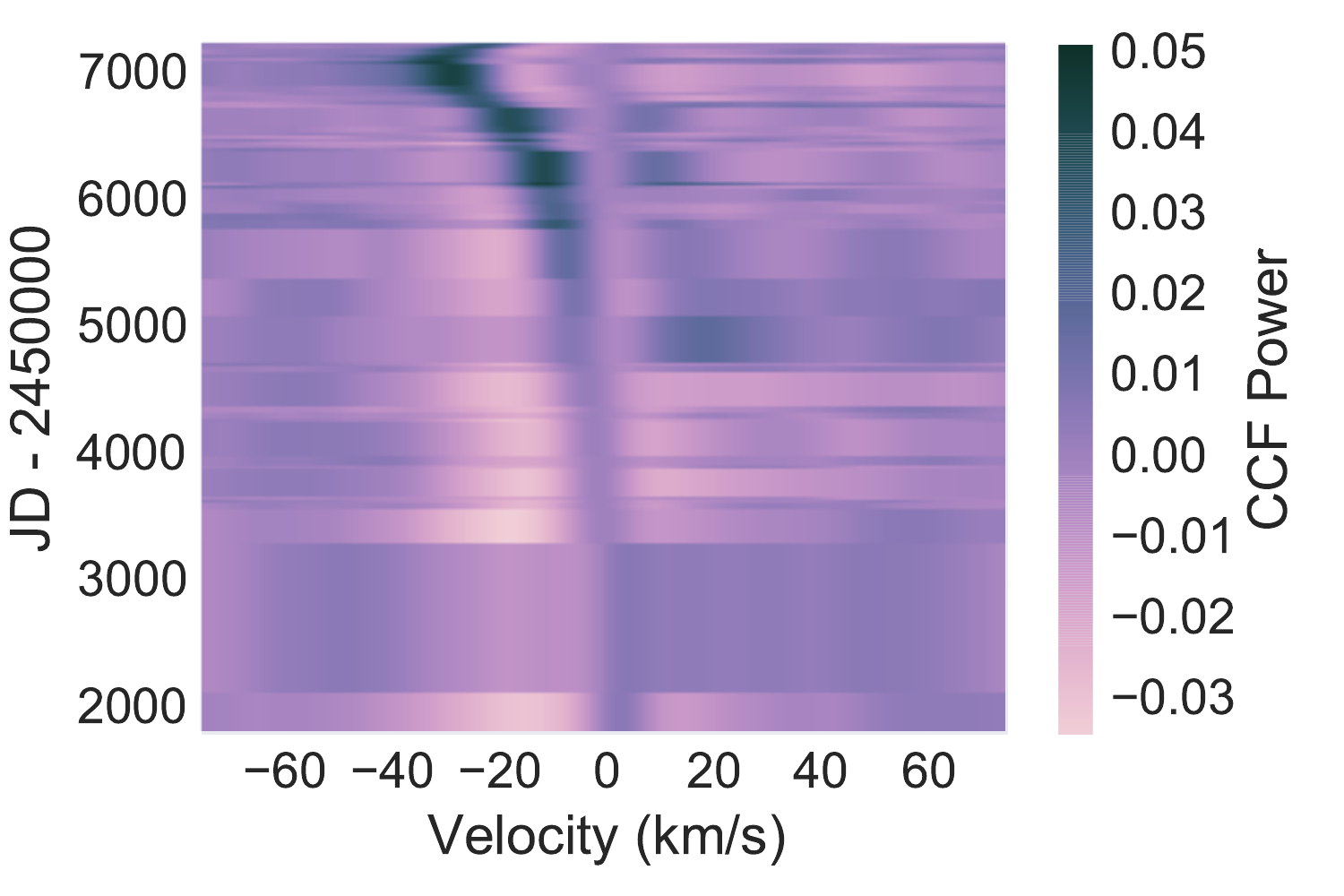}
  \caption{Cross-correlation functions of a 4400 K model spectrum template with the data, after subtraction of the average CCF. The dark curve in the top middle is the signal of the companion star.}
  \label{fig:resids}
\end{figure}

We normalize each CCF by subtracting a quadratic function that we fit well away from the peak, and then dividing by the height of the CCF at $v=0$ (the peak). The average of the shifted CCFs is a close estimate for the cross-correlation function of the M-star template with the F5 primary star, since the contribution from the companion is diluted by shifting the CCFs to the primary star rest frame. We remove the contribution from the primary star by subtracting the average from each CCF. The result is a series of residual cross-correlation functions that are estimates for the CCF of the companion spectrum against the 4400 K model spectrum template, with significant noise. We show the residual CCFs in Figure \ref{fig:resids}; the trace of the companion star is easily visible as the dark curve near the top middle. We are unable to recover the companion signal at early dates when the two stars were close to one another in velocity space.

\begin{figure}
  \centering
  \includegraphics[width=\columnwidth]{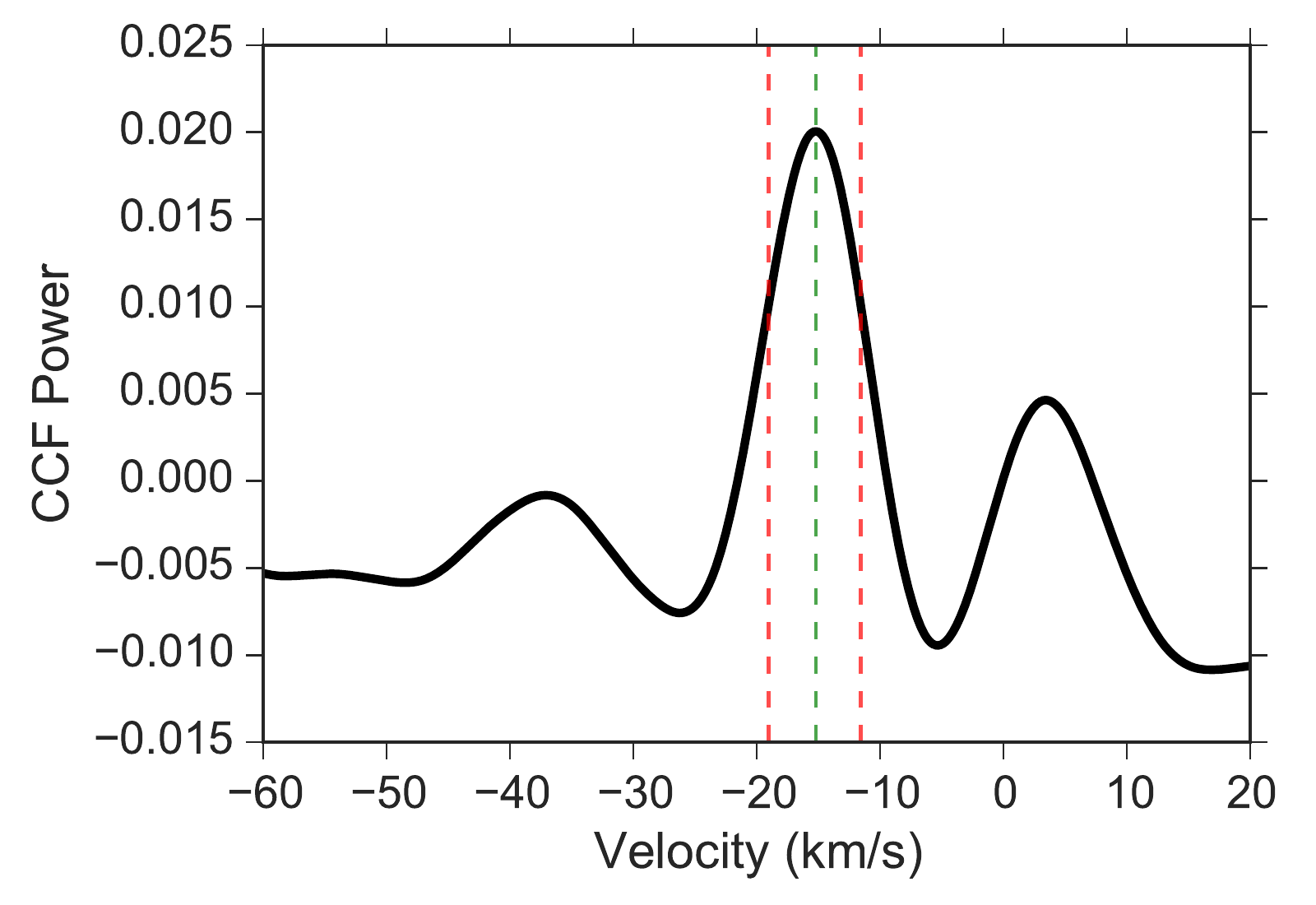}
  \caption{An example of a typical residual cross-correlation function. The dominant peak denotes the template match of the 4400 K template with the companion to $\psi^1$ Draconis A. The velocities are in the approximate rest frame of the primary star (see text for details), and the centroid and FWHM are given as vertical dotted lines.}
  \label{fig:ccf_typical}
\end{figure}

We measure the radial velocity of the companion at each epoch by finding the maximum and full-width-half-maximum (FWHM) of the residual CCF. On the basis of Figure \ref{fig:resids}, we use only the portion of the CCFs with $-50 < v < 10\ \rm km\ s^{-1}$; we show a typical residual CCF in Figure \ref{fig:ccf_typical}. Since the CCFs were shifted to subtract the contribution from the primary star, the measured velocities ($v_{m, 2}$) are related to the true barycentric velocities ($v_1$ and $v_2$ for the primary and secondary, respectively) and the constant shift described above ($\Delta v_2$) through

\begin{equation}
v_{m, 2}(t) = v_2(t) - v_1(t) + \Delta v_2
\label{eqn:vel}
\end{equation}

We give the measured companion velocities in Table \ref{tab:rv_data} (column 5). We additionally provide the velocities in the system velocity rest--frame ($v_2$) by using Equation \ref{eqn:vel} and the results of the analysis described below. The uncertainties given in Table \ref{tab:rv_data} are determined from the CCF peak width and the scaling factor ($f$) derived below. The shifted primary and secondary velocities given in Table \ref{tab:rv_data} are for the reader's convenience since they are in the same reference frame; we use the raw measurements in the orbital fit.

\section{Orbital Fit}
\label{sec:orbit}

We now use the radial--velocity measurements to find the best orbital parameters to describe the orbit, as well as some data scaling and shifting factors. The orbit is described by the semi-amplitudes for both the primary and secondary stars ($K_1$ and $K_2$, respectively), the longitude of pericenter ($\omega$), the eccentricity ($e$), the period ($P$), and the periastron--passage epoch ($T_0$). We cannot measure the system radial velocity, which is usually the final orbital element, because the measured primary--star velocities are differential and the secondary--star velocities are measured relative to the primary star.

Since the primary--star radial velocity measurements are differential measurements, we must also fit a constant shift ($\Delta v_1$) to account for the absolute radial velocity of the primary at the time at which our template spectrum was observed. We include an rv--jitter term ($\sigma_J$) to the fit to account for radial--velocity variations not encompassed by the orbital solution, and add the value in quadrature with the formal uncertainties on the primary star--velocity measurements. 

The companion radial velocities are measured relative to the primary star plus a small velocity shift ($\Delta v_2$) caused by slight inaccuracies in the vacuum--to--air wavelength conversion in the model spectrum and spectrograph wavelength drift throughout the night. Finally, the CCF peak full-width at half maximum vastly over-estimates the velocity uncertainty and so we fit a scale factor ($f$) to apply to the companion velocity uncertainties. The uncertainties given in Table \ref{tab:rv_data} are already scaled by that factor. The full log-likelihood function ($\mathcal{L}$) is then given by:

\begin{align*}
s_1 &= \sum_{t_m} \frac{(v_{1,m} - v_1(t_m) - \Delta v_1)^2 }{\sigma_{v_1}^2 + \sigma_J^2} + \ln{2\pi(\sigma_{v_1}^2 + \sigma_J^2)} \\
s_2 &= \sum_{t_m} \frac{(v_{2,m} - v_2(t_m)(1+\frac{K_1}{K_2}) - \Delta v_2)^2}{f\sigma_{v_2}^2 } + \ln{2\pi(f\sigma_{v_2}^2)} \\ 
\mathcal{L} &= -0.5(s_1 + s_2) \\
\end{align*}

where $v_{1,2}(t) = v(T_0, P, e, K_{1,2}, \omega, t)$ is the velocity at time t given by the orbital elements $T_0, P, e, K$, and $\omega$.

\begin{figure}

  \includegraphics[width=\columnwidth]{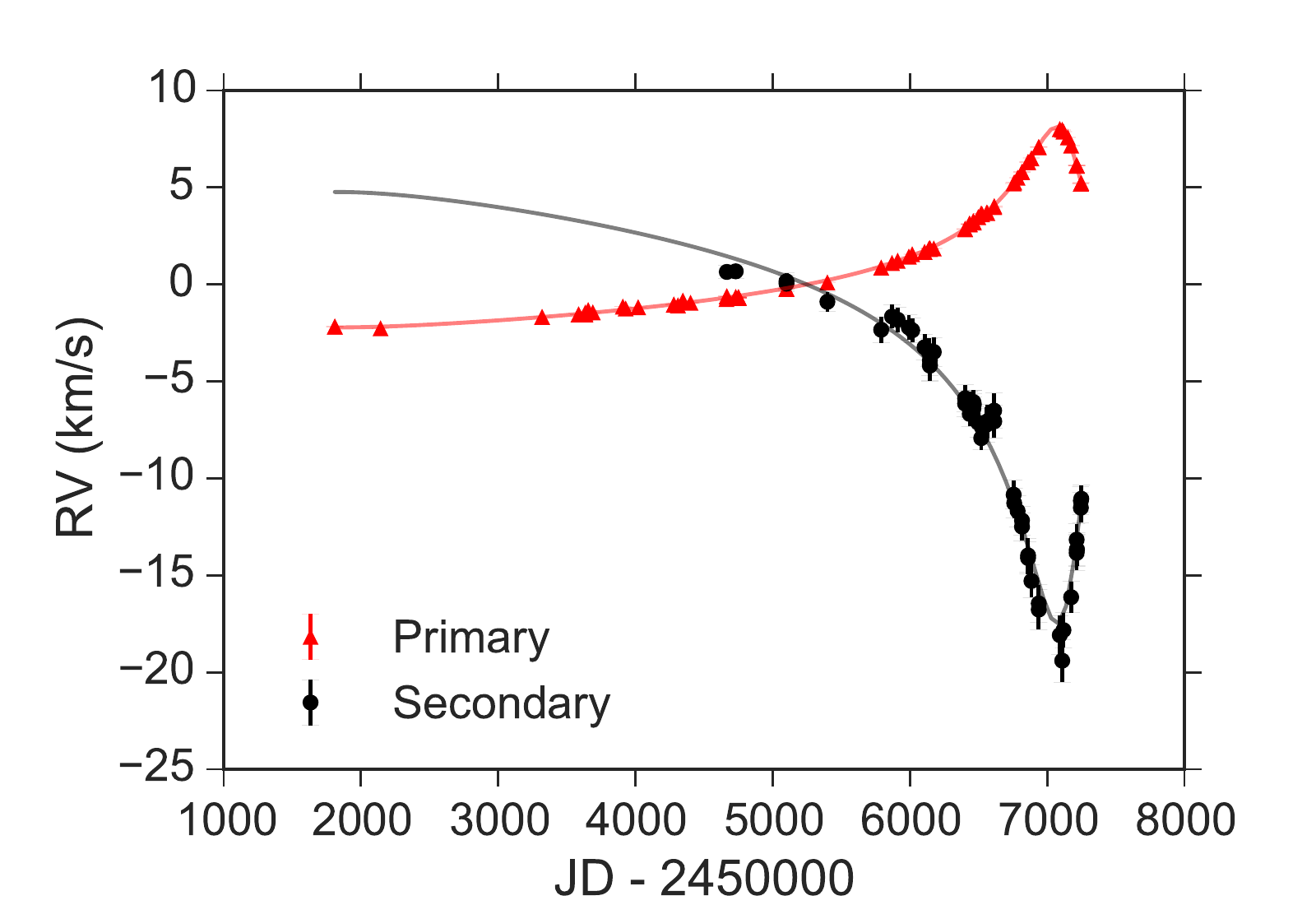}
  \caption{Best-fit double-lined orbit for the $\psi^1$ Draconis AC subsystem. There are no measurements of the companion at early dates because they could not be reliably measured in the residual cross-correlation functions.}
  \label{fig:orbit}
  
\end{figure}

We use the affine invariant sampler provided in the emcee code \citep{emcee} to perform a Markov Chain Monte Carlo (MCMC) fit to all of the parameters described above. We use flat priors in all variables except for the rv--jitter and companion rv uncertainty scale factors ($\sigma_J$ and $f$, respectively), for which we use log-uniform priors to allow for a large range of values.  We give the median value and uncertainty for each parameter in Table \ref{tab:orbit}. The uncertainties are estimated from the posterior probability distribution samples such that the lower and upper bounds give the 16th and 84th percentile (i.e. they are $1\sigma$ credibility intervals). We plot the best-fit orbit with the data in Figure \ref{fig:orbit}, with the uncertainties on the companion velocities scaled and the velocities shifted by $\Delta v_1$ and $\Delta v_2$. 

Next, we calculate a series of derived quantities to characterize the companion and report them in Table \ref{tab:orbit}. The mass ratio of the system is the ratio $K_1/K_2 = 0.47$. We estimate the primary star mass by interpolating Dartmouth isochrones \citep{Dotter2008} with the `isochrones' code \citep[described in][]{Montet2015}, and using spectroscopic parameters derived in \citet{Endl2015}. The secondary mass is $M_2 = qM_1 \sim 0.70\ M_{\odot}$; assuming the same age and metallicity as the primary, the Dartmouth isochrones give an expected temperature of ${\sim}4400 $ K. That temperature is in excellent agreement with the high--contrast--imaging data, which support a companion of ${\sim}4400$ K with large uncertainty. With both the primary and secondary star mass, we calculate the orbital inclination and semimajor axis and report them in Table \ref{tab:orbit}.

\begin{deluxetable}{rl}
\tabletypesize{\footnotesize}
\tablewidth{0pt}
\tablenum{2}
\tablecaption{ Orbital parameters for the $\psi^1$ Draconis A subsystem. }

\startdata
\cutinhead{Parameters from \citet{Endl2015}}
$T_{\rm eff,1}$ (K) & $6544 \pm 42$ \\
$\log{g}$ & $3.90 \pm 0.11$ \\
{[}Fe/H{]} & $-0.10 \pm 0.05$ \\

\cutinhead{Parameters derived in this work}
$K_1$ ($\rm km\ s^{-1}$) & $5.18^{+0.04}_{-0.03}$ \\
$K_2$ ($\rm km\ s^{-1}$) & $11.1 \pm 0.2$ \\
Period (days) & $6774^{+271}_{-167}$ \\
Periastron passage time (JD) & $2450388^{+169}_{-273}$ \\
$\omega$ (degrees) & $32.6 \pm 0.7$ \\
$e$ & $0.679^{+0.006}_{-0.004}$ \\
$\Delta v_1\ (\rm km\ s^{-1})$ & $4.10^{+0.06}_{-0.09}$ \\
$\Delta v_2\ (\rm km\ s^{-1})$ & $-5.4^{+0.3}_{-0.2}$ \\
Companion uncertainty scale factor ($f$) & $0.17 \pm 0.02 $\\
rv jitter (m s$^{-1}$) & $72^{+7}_{-5}$ \\
reduced $\chi^2$ & 0.41 \\ \\
$q$ & $0.466 \pm 0.008$ \\
$M_1$ (M$_{\odot}$) & $1.38^{+0.15}_{-0.08}$ \\
$M_2$ (M$_{\odot}$) & $0.70 \pm 0.07$ \\
$T_{\rm eff,2}$ (K) & $4400 \pm 300$ \\
$i$ (degrees) & $31 \pm 1$ \\
$a$ (AU) & $9.1^{+0.4}_{-0.3}$ \\

\enddata
\tablecomments{The primary mass is derived using the spectroscopic $T_{\rm eff}$, $\log{g}$, and [Fe/H] and interpolating Dartmouth isochrones. The companion temperature is likewise derived from the companion mass using Dartmouth isochrones of the same metallicity.}
\label{tab:orbit}
\end{deluxetable}

\section{Discussion and Conclusions}

We use nearly 15 years of time-series spectra of the star $\psi^1$ Draconis A to search for the spectral lines of a companion identified by a large--amplitude trend in the primary--star radial velocities and later by direct imaging. We cross-correlate each spectrum against a Phoenix model spectrum of a $4400$ K star and subtract the average CCF. The residual CCFs clearly show the template match with the companion (Figure \ref{fig:resids}), and we are able to measure the companion radial velocities for most dates. 

We use the radial--velocity measurements for both the primary and secondary stars to find an orbital solution for the now double-lined spectroscopic binary. The summary values of the fitted parameters are given in Table \ref{tab:orbit}. Finally, we report the mass and expected temperature of the companion as well as the orbital inclination and semi--major axis. The temperature agrees well with high--contrast imaging, validating our method.

The $\psi^1$ Draconis system is therefore a hierarchical multiple system with the component parameters given in Table \ref{tab:orbit}. $\psi^1$ Dra A and B are separated by ${\sim}680$ AU and have a mass-ratio $q = 0.82$, while A and C (the new companion) have a much closer orbit with with $a = 9.1$ AU and $q = 0.47$. 

This method could be used to search for the spectral lines of stellar companions to other stars observed with high--precision radial--velocity surveys. To that end, and in the goal of open science, we make the source code used for the analysis and generating the plots for this paper available at \url{https://github.com/kgullikson88/Companion-Finder}. The raw radial--velocity measurements for both the primary and secondary star, as well as the MCMC chains, are available at the same url.

This research has made use of the SIMBAD database, operated at CDS, Strasbourg, France, and of Astropy, a community-developed core Python package for Astronomy (Astropy Collaboration, 2013).
It was supported by a start-up grant to Adam Kraus from the University of Texas. The McDonald Observatory planet search is supported by the National Science Foundation under grant AST-1313075. We would like to thank the referee for various suggestions that improved this paper.

\newpage
\bibliography{}
\end{document}